\documentclass[a4paper,twocolumn,english,aps,prl]{revtex4}

\usepackage[T1]{fontenc}
\usepackage[latin1]{inputenc}
\usepackage{float}
\usepackage{graphicx,color}
\usepackage{bm}
\usepackage{bbm}
\usepackage{amsmath}
\usepackage{color}
\usepackage{amssymb}
\makeatletter

\usepackage{babel}
\makeatother
\begin{document}

\preprint{This line only printed with preprint option}

\title{Phase transitions in dipolar gases in optical lattices}

\author{Y. Sherkunov}

\affiliation{Physics Department, Lancaster University, Lancaster, LA1 4YB,
UK}
\author{Vadim V. Cheianov}

\affiliation{Physics Department, Lancaster University, Lancaster, LA1 4YB,
UK}

\author{Vladimir Fal'ko}

\affiliation{Physics Department, Lancaster University, Lancaster, LA1 4YB,
UK}

\begin{abstract}

We investigate the phase diagrams of two-dimensional lattice dipole systems with variable geometry.  For bipartite square and triangular lattices with tunable vertical sublattice separation, we find rich phase diagrams featuring a sequence  of easy-plane magnetically ordered phases separated by incommensurate spin-wave states. 

\end{abstract}
\maketitle

A recent breakthrough in the cooling of dipolar gases in optical lattices \cite{Chotia11}, following a decade of intensive research \cite{Weinstein98, Santos00, Micheli07,Ni08, Deiglmayr08, Aikawa09, Baranov08, Lahaye09}, opens a door into the earlier inaccessible many-body physics of lattice systems with anisotropic long-range interaction. Although bulk crystalline dipolar systems are abundant in nature \cite{Bell00, Lines}, their experimental investigation has been hindered by the extremely low temperatures required for  the observation of ordering transitions  \cite{Oja97} and the absence of continuously variable parameters. Recently, artificial two-dimensional dipolar systems, such as lithographically created nanomagnet arrays, have been realized \cite{Wang06, Moller06}. However, it is the advent of optical lattices with tunable lattice structure containing ultracold dipolar gases, that creates numerous possibilities for studies of previously unexplored phase transitions -- both classical and quantum -- between  ordered and disordered  phases of this fundamental many-body system. 

\begin{figure}[ht]
\centering
 \includegraphics[width=0.48\textwidth]{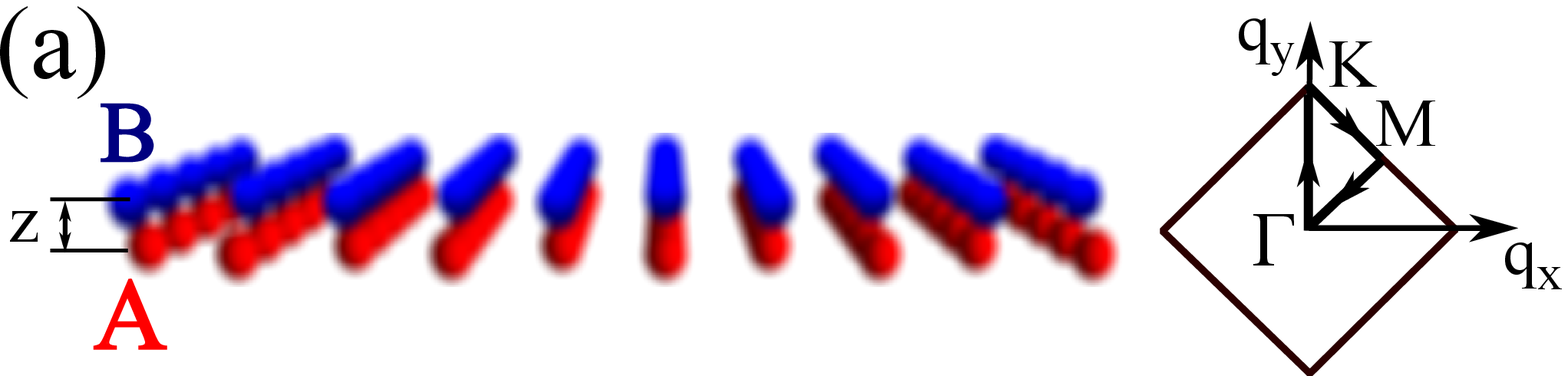}
  \includegraphics[width=0.47\textwidth]{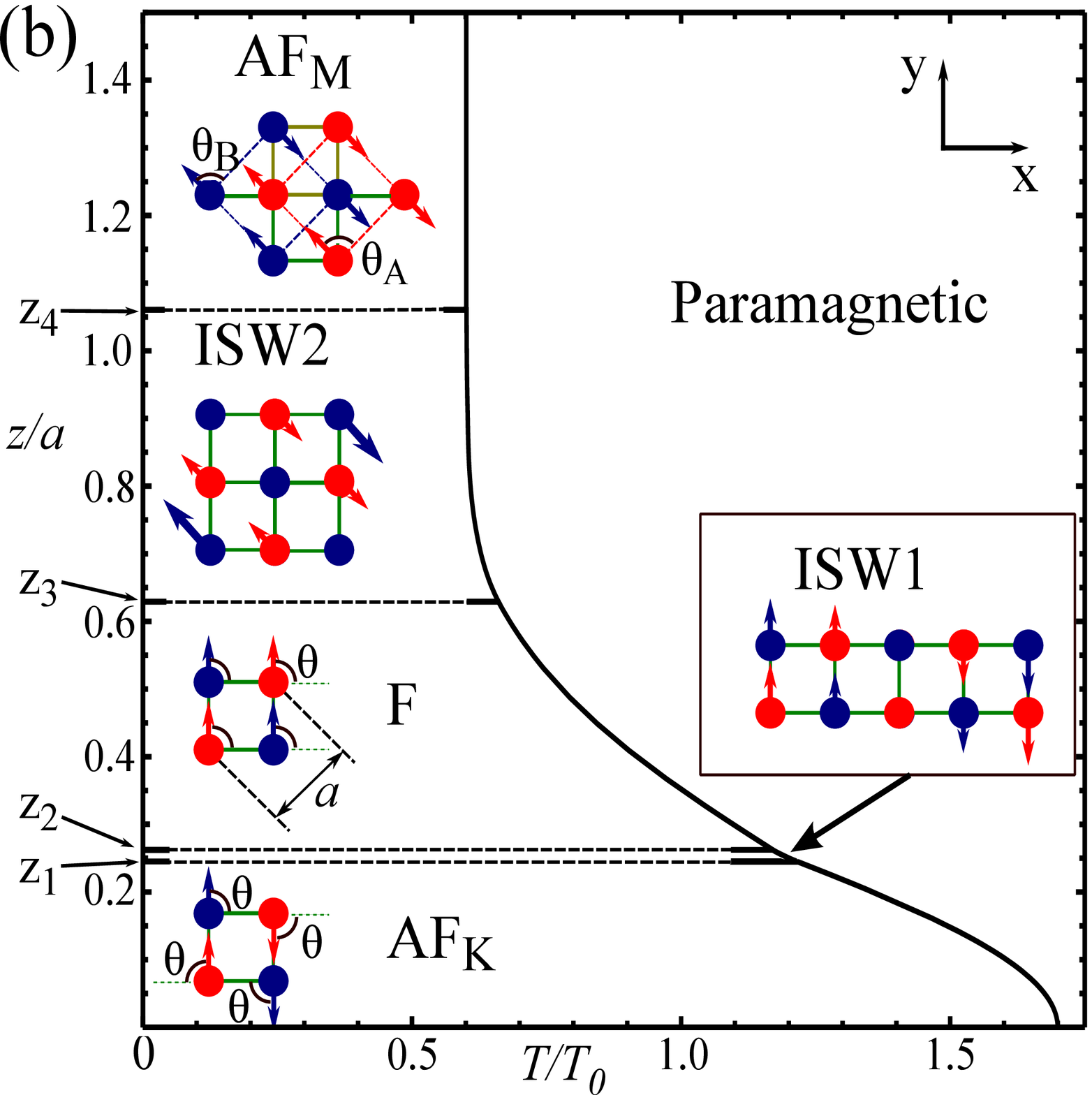}
\caption{\label{fig:fig1} (color online) (a)  Dipolar magnetic gas on a bipartite square  optical lattice as seen along [0,1] axis and the first Brillouin zone. The two sublattices, $A$ and $B$, are vertically separated  by the distance $z$. (b) Phase diagram of the dipolar gas:  three commensurate phases ($\mathrm{AF_K}$, F, and $\mathrm{AF_M}$) separated by two incommensurate spin-wave phases (ISW1 and ISW2) with phase boundaries at   $z_{1}=0.245a$, $z_{2}=0.262a$, $z_{3}=0.629a$ and $z_{4}=1.061a$, where $a$ is lattice constant, and $T_0=\mu_0\mu^2/4\pi a^3$. }
\end{figure}

\begin{figure}
\centering
 \includegraphics[width=0.48\textwidth]{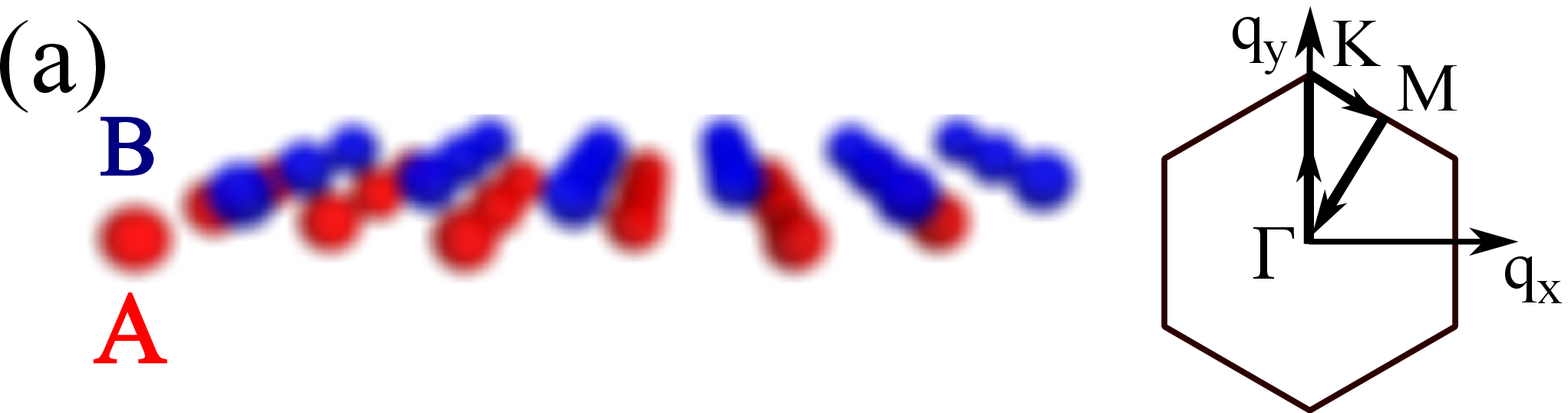}
  \includegraphics[width=0.48\textwidth]{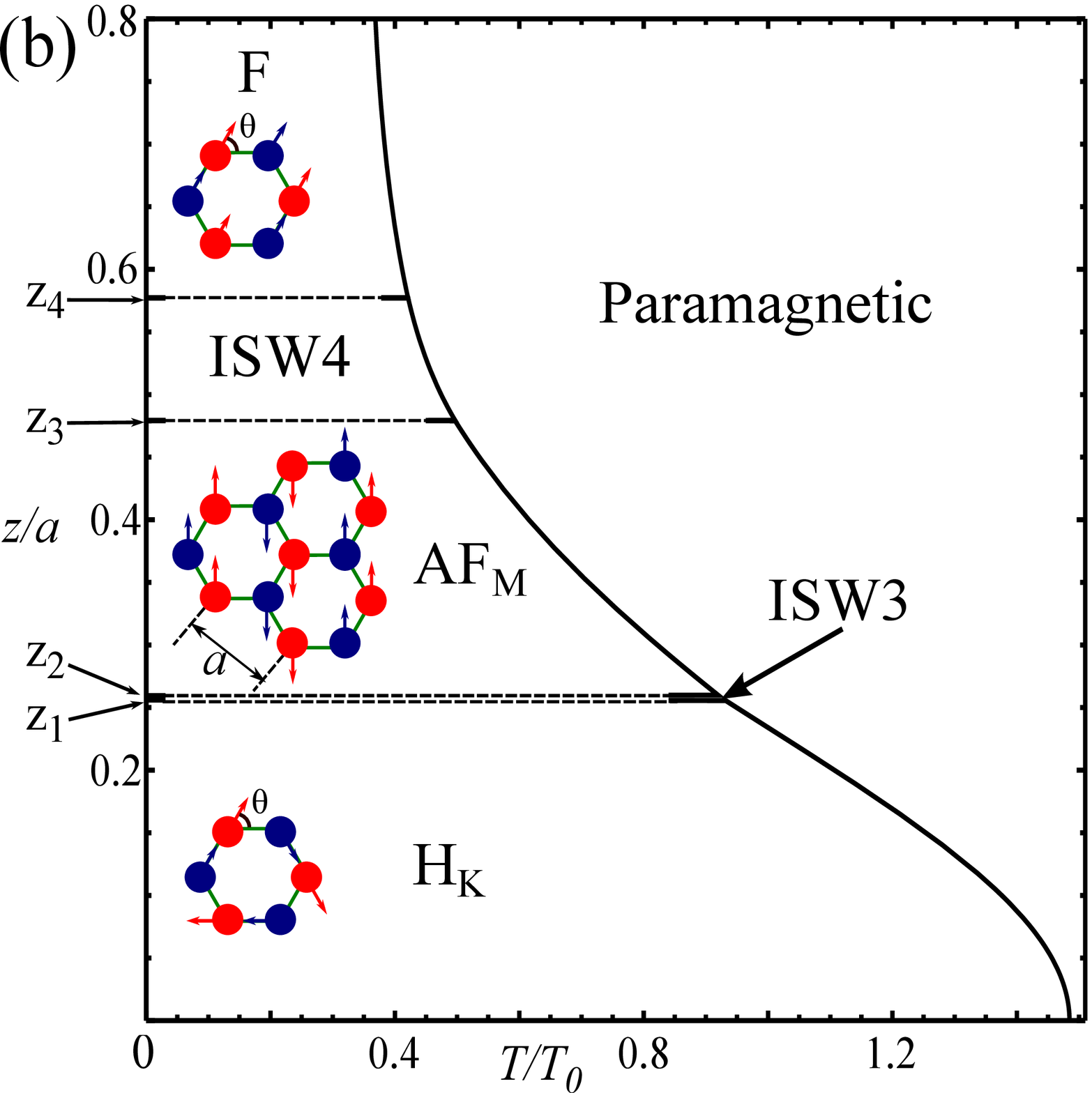}
\caption{\label{fig:fig2} (a) Dipolar magnetic gas on a bipartite triangular optical lattice as seen along [0,1] axis and the first Brillouin zone.  The two sublattices, A and B, are vertically  shifted separated  by the distance $z$  (at $z=0$, they form a honeycomb lattice). (b) Phase diagram of the dipolar gas:  three commensurate phases ($\mathrm{H_K}$, $\mathrm{AF_M}$, and F) separated by two incommesurate spin-wave phases (ISW3 and ISW4) with phase boundaries at  $z_{1}=0.256a$, $z_{2}=0.2600a$, $z_{3}=0.479a$ and $z_{c4}=0.577a$, where $a$ is  lattice constant and $T_0=\mu_0\mu^2/4\pi a^3$.     }
\end{figure}

In this Article we analyze a series of  magnetic phase transitions in a  classical  dipolar gas in deep optical lattices  [square, Fig. \ref{fig:fig1} and triangular, Fig. \ref{fig:fig2}] obtained from bipartite monolayer lattices by  vertical separation, $z$, of the two sublattices. One way to realize such  systems would be loading of ferromagnetic spinor Bose-Einstein mini-condensates in the nodes \cite{bluedetuned} of  a deep bilayer optical lattice created with the help of a painted potential technique \cite{Henderson09}, which would allow for high degree of control over the shapes of optical lattices and interlayer separation. 

 We find that, upon the variation of  $z$, each system experiences a sequence of easy-plane magnetically ordered phases separated by incommensurate spin-wave states,  which could be detected with the help of Bragg diffraction of light \cite{Weidemuller95, Birkl95,Corcovilos10}.  The phase diagram for the square lattice on the $z-T$ plane  is shown in Fig. \ref{fig:fig1}. For sufficiently small separations $z\ll a$, where $a$ is lattice constant, we reproduce the earlier predicted \cite{Rozenbaum96,Gross02} canted antiferromagnetic phase, $\mathrm{AF_K}$, with the ordering vector $\mathbf K$.  For $z>a$, we find an antiferromagnetic phase, $\mathrm{AF_M}$, with larger unit cell and ordering wave vector at the M-point of the Brillouin zone of the bipartite lattice.   For intermediate interlayer distances, we find a stable ferromagnetic phase (F), separated from the antiferromagnetic ones by incommensurate spin-wave states (ISW). At the critical temperature $T_c$, all of the ordered phases feature a degeneracy in the orientation of magnetization, characterized in  Fig. \ref{fig:fig1} by angle $\theta$, or $\theta_A$ and $\theta_B$ for $\mathrm{AF_M}$. 
Away from $T_c$, such a degeneracy is lifted, and  Fig. \ref{fig:fig1} shows the optimal orientation of the order parameter for the low-temperature states. The structure of the intermediate incommensurate phases, ISW1 and ISW2 (Fig. \ref{fig:fig1}) has also been established for $T\rightarrow T_c$, while their nature at low temperatures remains an open question.  

We find that the phase diagram for the bipartite triangular lattice (which forms a honeycomb lattice when $z=0$) also contains a sequence of commensurate and incommensurate magnetic phases, Fig.\ref{fig:fig2}. For $z\ll a$, we find a helical phase with the ordering vector $\mathbf K$ ($\mathrm {H_K}$), which is specific for dipoles on a  2D honeycomb lattice \cite{Rozenbaum96} . A large vertical displacement of the two sublattices of the honeycomb lattice results in two weakly coupled triangular lattices, for which the ground state is ferromagnetic (F) \cite{Rozenbaum96}. In between those two extremes lies an antiferromagnetic phase $\mathrm{AF_M}$ with the ordering vector $\mathbf{M}$ \cite{Mphase},  separated from the helical and ferromagnetic phases by parametric intervals, where the magnetization texture is incommensurate with the lattice.  
 
To find  phase diagrams in Figs. \ref{fig:fig1}, \ref{fig:fig2},  we consider an ensemble of classical magnetic dipoles $\mu\mathbf s_i$ ($|\mathbf s_i|^2=1$), placed on the A or B sites  $\mathbf{r}_i^{n}$ ($n=A,B$) of the square (Fig. \ref{fig:fig1}) or triangular (Fig. \ref{fig:fig2}) lattices, with interaction energy 
\begin{alignat}{2}
H=\frac{1}{2}\sum_{i\neq j} \sum_{\alpha, \beta =x,y,z} &\sum_{n,n'=A,B}J^{\alpha \beta}(\mathbf r_{i}^{n}-\mathbf r_j^{n'})s^{\alpha}(\mathbf r_i^{n}) s^{\beta}(\mathbf r_j^{n'})\nonumber,\\
J^{\alpha \beta}(\mathbf{r})&=\frac{\mu_0\mu^2}{4\pi r^3}\left(\delta_{\alpha \beta}-3\frac{r^{\alpha}r^{\beta}}{r^2}\right) \label{J}.
\end{alignat}
Here $\mu_0$ is the vacuum permeability. The Hamiltonian is invariant under the group $\mathcal{G}$ of simultaneous rotation of magnetic moments in the $xy$-plane and the lattice rotations through the angles $\pi/4$ for square and $\pi/3$ for triangular lattices. 

In order to identify the thermodynamic average, $\langle \mathbf s(\mathbf r_i^n)\rangle$ ($n=A,B$), of magnetization for various interlayer separations $z$, we apply the Landau theory  and study the free energy in the vicinity of the transition temperature \cite{Palmer00},
\begin{eqnarray}
F=\frac{N}{2}\sum_{\mathbf q}\mathbf{S}^{\dagger}({\mathbf q )}\left(3T\hat{\mathbbm{1}}+\hat{\mathrm{J}}(\mathbf q)\right)\mathbf{S}(\mathbf{q})+F^{(m)}+...,
\label{F}
\end{eqnarray}
expressed in terms of a 6-vector
\begin{eqnarray}
\mathbf{S(\mathbf q)}^{\mathrm{T}}=(S_A^x(\mathbf q),S_A^y(\mathbf q),S_A^z(\mathbf q),S_B^x(\mathbf q),S_B^y(\mathbf q),S_B^z(\mathbf q)),\nonumber
\end{eqnarray}
where $S_{n}^{\alpha}(\mathbf q)$
 is the Fourier transform of the order parameter $\langle \mathbf s(\mathbf r_i^n)\rangle$ of magnetic moments  on A and B sublattices. In Eq. \ref{F}, $\hat{\mathbbm 1}$ is a $6\times 6$ unit matrix, and   $F^{(m)}$ incorporates higher-order invariants under the group $\mathcal{G}$ built using the order parameter ($m=4$ for square and $m=6$ for triangular lattices).
A $6\times 6$ matrix $\hat{\mathrm J}$ has elements
\begin{eqnarray}
 J_{nn'}^{\alpha\beta}(\mathbf{q})=(1/N)\sum_{ij}J^{\alpha\beta}(\mathbf r_{i}^{n}-\mathbf r_{j}^{n'})e^{i\mathbf q\cdot (\mathbf r_{i}^{n}-\mathbf r_{j}^{n'})}, 
 \label{J}
 \end{eqnarray}
 where $N$ is the number of unit cells. For each wave vector $\mathbf q$, matrix $\hat{\mathrm{J}}$ has $6$ eigenvalues, $\lambda$, and $6$ eigenvectors, $\mathbf V(\mathbf q)$. The lowest negative eigenvalue $\lambda_0(\mathbf q_0)$ found among $\lambda_{\gamma}(\mathbf q)$ by varying wave vector $\mathbf q$ over the Brillouin zone   determines the polarizations $\mathbf V(\mathbf q_0)$ and the wave vector $\mathbf q_0$ of the most favorable magnetic ordering and  the transition  temperature
\begin{eqnarray}
T_{c}=-\frac{1}{3}\min_{\gamma,\mathbf q} (\lambda_{\gamma}(\mathbf q))\equiv\frac{1}{3}|\lambda_0(\mathbf q_0)|.
\label{T}
\end{eqnarray}

\begin{figure}[!]
\centering
 \includegraphics[width=0.48\textwidth]{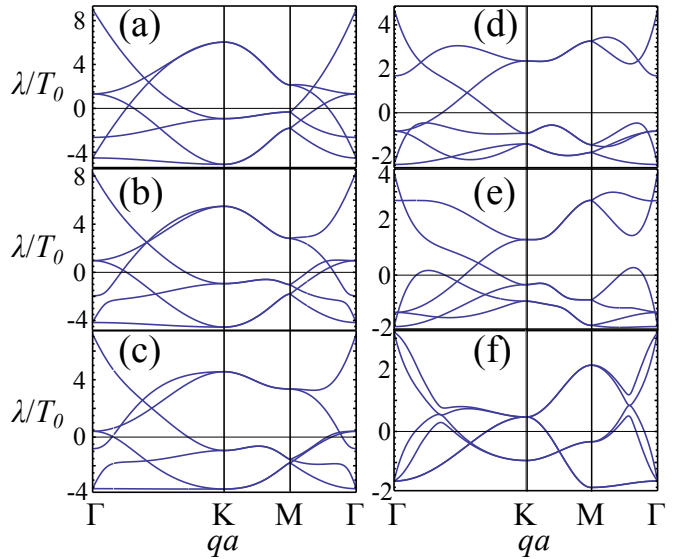}
\caption{\label{fig:fig3}  Eigenvalues, $\lambda_i(\mathbf{q})$, of the dipolar tensor, $J_{nn'}^{\alpha \beta}$,  for the square lattice as  functions of the wave vector  $\mathbf q$ along symmetric directions in the Brillouin zone (see  Fig.\ref{fig:fig1}(a)) for a set of lattice displacements, $z$: (a) $z=0$; (b) $z=0.140a$; (c) $z=0.248a$; (d) $z=0.500a$; (e) $z=0.707a$; (f) $z=2.120a$. }
\end{figure}

\begin{figure}[!]
\centering
 \includegraphics[width=0.48\textwidth]{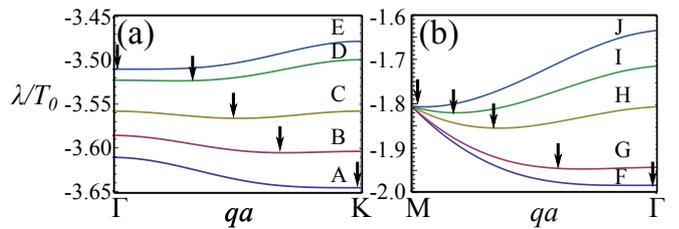}
\caption{\label{fig:fig4} The minimal eigenvalues $\lambda_0$ vs. $\mathbf q$ along symmetric directions in the Brillouin zone for incommensurate phases: (a) ISW1  and (b) ISW2 for a representative set of $z$:  (A) $z=0.245a$; (B) $z=0.250a$; (C) $z=0.255a$; (D) $z=0.260a$; (E) $z=0.262a$; (F) $z=0.629a$; (G) $z=0.651a$; (H) $z=0.743a$; (I) $z=0.849a$; (J) $z=1.061a$. Arrows show the positions of the minima, $\mathbf q_0$, of $\lambda$. The ordering vector, $\mathbf q_0$ continuously moves along the straight lines connecting the K- and $\Gamma$- points (a) and  $\Gamma$- and M-points (b), as $z$ increases.}
\end{figure}

In  Fig. \ref{fig:fig3}, we show plots for  $\lambda_{\gamma}(\mathbf q)$ for the square bipartite lattice with various vertical A-B sublattice separations. For  $0\leq z\leq z_{1}$, where $z_{1}=0.245a$ ($a$ is lattice constant) $\mathbf q_0$ coincides with  one of  $\mathbf K$-points of the Brillouin zone as shown in Fig.\ref{fig:fig3} (a)-(b). This corresponds to the $\mathrm{AF_K}$ phase (Fig. \ref{fig:fig1}) with the order parameter  
\begin{eqnarray}
\langle \mathbf s(\mathbf r_i^{n})\rangle \propto(\cos(\mathbf K\cdot \mathbf r_i^{n}+\theta), p \sin(\mathbf K\cdot \mathbf r_i^{n}+\theta)),
\label{K} 
\end{eqnarray}
where $p=\pm 1$ for $n=A/B$, and $\theta$ is a constant.  In Eq. (\ref{F}), the degeneracy in  $\theta$  is  lifted by the higher-order  terms $F^{(m)}$ appearing after taking into account thermal fluctuations. As $z$ increases from $z_1$ to $z_2=0.262a$, $\mathbf q_0$  continuously moves from  $\mathbf K$- to   $\Gamma$- point (Fig.\ref{fig:fig4} (a)),  and the corresponding eigenvector $\mathbf V_0(\mathbf q_0)$ determines the magnetization texture
\begin{eqnarray}
\langle \mathbf{s}(\mathbf{r}_i^{n})\rangle\propto \hat{\mathbf{z}}\times \mathbf{q}_0\cos(\mathbf{q}_0\cdot \mathbf{r}_i^{n})
\label{B}
\end{eqnarray}
of the incommensurate phase ISW1 illustrated in Fig. \ref{fig:fig1} (b), 
 where $\hat z$ is the unit vector perpendicular to the plane of the lattice and $\mathbf q_0=(q_0,0)$. For $z_2\leq z\leq z_3$, where $z_3=0.629a$, $\mathbf q_0$ lies at $\Gamma$-point (Fig. \ref{fig:fig3} (d)), which corresponds to the easy-plane ferromagnetic (F)  ordering.  As $z$ increases from $z_3$ to $z_4=1.061a$,  $\mathbf q_0$ continuously moves from  $\Gamma$-  to  $M$-point (Fig. \ref{fig:fig4} (b)), which determines the incommensurate spin-wave state ISW2 and the order parameter given by Eq. (\ref{B}) with $\mathbf q_0=(1/\sqrt 2)(q_0,q_0)$. For $z\geq z_4$,  $\mathbf q_0$ is at one of  $M$-points (Fig. \ref{fig:fig3} (f)) giving rise to the phase $\mathrm{AF_M}$, which can be viewed as two weakly coupled "$\mathrm{AF_K}$" phases on each of two square sublattice. The form of the order parameter in each of the commensurate phases is given in Table \ref{table1}.

The phase diagram for the bipartite triangular lattice (Fig. \ref{fig:fig2} (b) with order parameters listed in Table \ref{table1}) is somewhat similar to that for the square bipartite lattice.  For $0\leq z\leq z_1$, where $z_1=0.256a$, we find that $\mathbf q_0$  is at one of the  $\mathbf{K}$-points. This corresponds to the helical $\mathrm{H_K}$ phase with the order parameter given by Eq. (\ref{K}), where vector $\mathbf K$ is at the corner of the hexagonal Brillouin zone of triangular lattice (Fig. \ref{fig:fig2} (a)). Such a phase has been discussed in relation with a dipolar gas in a planar honeycomb lattice \cite{Rozenbaum96}. For $z_1\leq z \leq z_2$, where $z_2=0.260a$,  $\mathbf q_0$  continuously shifts from $\mathbf K$ to $\mathbf M$-point giving rise to the incommensurate phase ISW3 with magnetization texture
\begin{eqnarray}
\langle \mathbf s(\mathbf r_i^{n})\rangle\propto(-\sin(\mathbf q_0\cdot \mathbf r_i^{n}), pa_y \cos(\mathbf q_0\cdot \mathbf r_i^{n}),a_z\sin(\mathbf{q}_0\cdot \mathbf r_i^{n})),\nonumber
\end{eqnarray}
where   $0\leq a_y$ and $a_z\ll a_y$ are z-dependent, $\mathbf q_0=q_0(\mathbf{K}(1-c)+\mathbf M c)/|(\mathbf{K}(1-c)+\mathbf M c)|$ ($0\leq c \leq 1$). For $z_2\leq z \leq z_3$, where $z_3=0.479$,  $\mathbf q_0$ lies at one of the M-points, which corresponds to the easy-plane antiferromagnetic phase $\mathrm{AF_M}$ shown in Fig. \ref{fig:fig2} (b) \cite{Mphase}.    For $z_3\leq z \leq z_4$, where $z_4=0.577a$, $\mathbf q_0$  moves from $\mathbf{M}$- to $\Gamma$-points, which determines the incommensurate spin-wave state ISW4,
\begin{equation}
\langle \mathbf s(\mathbf r_i^{n})\rangle\propto(\sin(\mathbf q_0\cdot \mathbf r_i^n),\cos(\mathbf q_0\cdot \mathbf r_i^n))
\end{equation}
with $\mathbf q_0=q_0\mathbf M/|\mathbf M|$.
Finally, for $z\geq z_4$,  $\mathbf q_0$  lies at  $\Gamma$-point, which corresponds to an easy-plane ferromagnetic state.  In the limit $z\rightarrow\infty$, this coincides with the ground state calculated for a dipolar magnet on a plane  triangular lattice \cite{Rozenbaum96}.  

\begin{table*}[ht]
\hfill{}
\caption{Order parameter for each of  the commensurate phases of a dipolar gas on bipartite square and triangular lattices.}  % title of Table
\centering % used for centering table
\begin{tabular}{|c| c| c| c| c| } % centered columns (4 columns)
\hline\hline %inserts double horizontal lines
Lattice & Phase&  $\mathbf S\equiv \langle \mathbf{s}(\mathbf{r}_{i}^{n})\rangle/|\langle \mathbf{s}(\mathbf{r}_{i}^{n})\rangle|$ & $\mathbf q_0$ &$\theta_0$ \\ [0.5ex]
%heading
\hline % inserts single horizontal line
Square & $\mathrm{AF_K}$ &$(\cos (\mathbf K\cdot \mathbf r_i^{n}+\theta),p \sin (\mathbf K\cdot \mathbf r_i^{n}+\theta))$,  $p=\pm 1$ for A/B & $\mathbf{K}=(0,\frac{\sqrt 2\pi} {a})$ & $\frac{\pi k}{2}$, $k\in \mathbb{Z}$\\[1ex]  \cline{2-5}
$n=A,B$& F &$(\cos \theta ,\sin \theta)$&$0$&$\frac{\pi k}{2}$, $k\in \mathbb{Z}$ \\[1ex]\cline{2-5}
& $\mathrm{AF_M}$ &$(\cos(\mathbf M\cdot \mathbf r_i^n\pm \theta_{n}+\pi/4),\sin(\mathbf M\cdot \mathbf r_i^n\pm\theta_n+\pi/4))$ \cite{AB}&$\mathbf M=(\frac{\sqrt 2\pi}{ 2a},\frac{\sqrt 2\pi}{2a})$&$\frac{\pi k}{2}$, $k\in \mathbb{Z}$ \\[0.1ex]
\hline 
Triangular  & $\mathrm{H_K} $&$(\cos (\mathbf K\cdot \mathbf r_i^{n}+\theta),p \sin (\mathbf K\cdot \mathbf r_i^{n}+\theta))$&$\mathbf K=(0,\frac{4\sqrt 3\pi}{3a})$ &$\frac{\pi k}{3}$,  $k\in \mathbb{Z}$\\ [1ex] \cline{2-5}% [1ex] adds vertical space\
$n=A,B$ &  $\mathrm{AF_M}$ &$(\sin (\mathbf M\cdot \mathbf r_i^{n}),\cos (\mathbf M\cdot \mathbf r_i^{n}))$&$\mathbf M=(\frac{2\pi}{\sqrt 3a},0)$ &---\\[1ex]\cline{2-5}
  &  F &$(\cos \theta,\sin \theta)$&$0$ &$\frac{\pi k}{3}$,  $k\in \mathbb{Z}$\\[1ex]\cline{2-5}
   \hline %inserts single line
\end{tabular}
\label{table1} % is used to refer this table in the text
\end{table*}

The above analysis of magnetic phases of dipolar gases on square and triangular bipartite lattices, limited to the quadratic terms in the Landau theory, is formally valid at  $T\rightarrow T_c$.  To extend the phase diagrams in Figs. \ref{fig:fig1} and \ref{fig:fig2} to low temperatures, we investigate the stability of the ordering patterns described in Table \ref{table1} near $T=0$ using the linear spin-wave theory \cite{Palmer00}. That is, we expand the interaction energy  
 in small deviations of on-site magnetic moments from the ground-state value  $\langle\mathbf s(\mathbf r_i^k)\rangle$,
 \begin{eqnarray}
E=E(\mathbf S)+\frac{N_M}{2}\sum_{\mathbf q}\mathbf {\sigma}^{\dagger}(\mathbf q) \hat{\mathrm A}(\mathbf q) \mathbf {\sigma}(\mathbf q).
\label{flucf}
\end{eqnarray}
 \noindent Such deviations have to respect the constraint $|\mathbf s(\mathbf r_i^k)|^2=1$ and can be parametrized as 
 \begin{align}
 \mathbf s(\mathbf r_i^k) = \langle\mathbf  s(\mathbf r_i^k)\rangle+\hat {\mathbf z}\sigma_{z}(\mathbf r_i^k)+(\hat {\mathbf{z}}\times \langle\mathbf s(\mathbf r_i^k))\rangle\sigma_{||}(\mathbf r_i^k)\nonumber\\
 -(1/2)\langle\mathbf s(\mathbf r_i^k)\rangle[(\sigma_{||}(\mathbf r_i^k))^2+(\sigma_{z}(\mathbf r_i^k))^2]. \nonumber
 \end{align}
\noindent Here we use the Fourier transform, $\mathbf {\sigma} (\mathbf q)^{\mathrm{T}}= ( \sigma^1_{||}(\mathbf q), \sigma^1_{z} (\mathbf q),...,  \sigma^k_{||}(\mathbf q), \sigma^k_{z} (\mathbf q),...)$ of   $\sigma_{\alpha}(\mathbf r_i^k)$, index $k=1,...,M$  labels sites within the magnetic unit cell of a commensurate phase, so that  the Fourier transform of $\langle\mathbf s(\mathbf r_i^k)\rangle$ is $\mathbf S_k(\mathbf q)=\mathbf S_k\delta (\mathbf q)$ and $N_M$ is the number of magnetic unit cells.  The $2M\times 2M$ matrix $\hat{\mathrm A}(\mathbf q)$ has elements $A_{kk'}^{||,||}(\mathbf q)=\sum_{\alpha\beta}(\hat{\mathbf z}\times \mathbf S_k)^{\alpha}J_{kk'}^{\alpha\beta}(\mathbf q)(\hat{\mathbf z}\times \mathbf S_{k'})^{\beta}+f(0)$, $A_{kk'}^{||,z}(\mathbf q)=\sum_{\alpha}(\hat{\mathbf z}\times \mathbf S_k)^{\alpha}J_{kk'}^{\alpha z}(\mathbf q)$, $A_{kk'}^{z,||}(\mathbf q)=\sum_{\beta}J_{kk'}^{z\beta}(\mathbf q)(\hat{\mathbf z}\times \mathbf S_{k'})^{\beta}$, $A_{kk'}^{zz}(\mathbf q)=J_{kk'}^{zz}(\mathbf q)+f(0)$, where $f(0)=(-1/2)\sum_{\alpha\beta}S_k^{\alpha}J_{kk'}^{\alpha\beta}(0)S_{k'}^{\beta}$.  We find that  all eigenvalues of $\hat{\mathrm A}(\mathbf q)$ are positive for $\mathbf q\neq 0$ within the same intervals $z_i<z<z_{i+1}$ , whereas at the edges of the intervals, the lowest eigenvalue of $\hat{\mathrm A}(\mathbf q)$ acquires negative values reflecting a tendency towards incommensurate textures.  
 
 For most of the phases in Table \ref{table1},  the interaction is degenerate in angle $\theta$ (Figs. \ref{fig:fig1} and \ref{fig:fig2}). This degeneracy is  lifted \cite{Prakash}  by thermal fluctuations leading to the higher-order expansion terms in the Landau theory.  To find, at least for $T\ll T_c$, such symmetry  breaking contributions, we take into account fluctuations  $\mathbf{\sigma}(\mathbf q)$ of the order parameter following \cite{Prakash} and calculate the entropy part of the free energy 
\begin{eqnarray}
\delta F_s=-T\ln\prod_{\mathbf{q}}\int D\mathbf {\sigma}(\mathbf q) \exp [-\frac{1}{2T} \mathbf {\sigma}^{\dagger}(\mathbf q) \hat{\mathrm A}(\mathbf q) \mathbf {\sigma}(\mathbf q)]\nonumber\\
\approx \frac{N_MT}{2}\sum_{\mathbf{q}\in BZ}\ln(\det \hat{\mathrm{A}}(\mathbf q)).
\label{Z}
\end{eqnarray}
We evaluate the dominant contribution from the fluctuations lifting the degeneracy with respect to angle $\theta$:
\begin{eqnarray}
\delta F_s\approx TA_0-TA_1\cos (m\theta), A_0,A_1>0,
\end{eqnarray}
where $m=4$ for square and $m=6$ for triangular lattice \cite{AB}. This determines the optimal choice $\theta_0$ shown in Table \ref{table1}. For $T\rightarrow T_c$, such entropy terms give rise to the crystalline anisotropy contribution $F^{(m)}\propto T_c|\mathbf S|^m\cos m\theta$ in Eq. (\ref{F}).    

For magnetic dipolar gases in deep bipartite (bilayer) square and triangular optical lattices, the predicted phase diagram may appear very much within the experimentally accessible range of controlled parameters. For deep optical lattices with $a\sim 1\mathrm{\mu m}$ and optical field trapping mini-condensates  of $10^3$ spin-aligned $^{87}\mathrm{Rb}$ atoms per unit cell, we estimate $T_c\sim 50 \mathrm{\mu K}$ in the phase diagram in Figs. \ref{fig:fig1} and \ref{fig:fig2}.  Moreover, as the electric and magnetic dipole interactions are mathematically equivalent, the phase diagram in Figs. \ref{fig:fig1} and \ref{fig:fig2} should be applicable to the electric dipolar systems, where we estimate  $T_c \sim 100\mathrm{nK}$ for ferro- and antiferroelectric transitions in molecules with a dipole moment $d\sim 1\mathrm D$.

%\begin{figure}[t] 
%\centering
 %\includegraphics[width=0.48\textwidth]{figpdtr}
  %\caption{\label{fig:fig8}        }
%\end{figure}

\bibliographystyle{apsrev}
\bibliography{c13}

\begin{thebibliography}{27}
\bibitem[{blu()}]{bluedetuned}
\bibinfo{note}{The presence of the trapping potential can have an important
  effect on the ordering patterns discussed here because of light-induced
  dipole-dipole interaction. However, this effect can be neglected for
  blue-detuned lattices, where atoms are located in the nodes \cite{Zhang02}.}
\bibitem[{Mph()}]{Mphase}
\bibinfo{note}{Similar phase has been recently predicted in \cite{Varney} for
  an XY- spin model on a honeycomb lattice with next-to-nearest neighbors
  interactiion}.
\bibitem[{AB()}]{AB}
\bibinfo{note}{For the $\mathrm{AF_M}$ phase of the dipolar gas in the square
  lattice the interaction energy is degenerate in two parameters, $\theta_A$
  and $\theta_B$ (see Fig. \ref{fig:fig1}). This degeneracy is lifted by
  thermal fluctuations leading to $\delta F_s\approx TA_0-TA_1[\cos
  (m\theta_A)+\cos (m\theta_B)]$, and the anisotropy term of the Landau theory
  is $F^{(m)}\propto T_c|\mathbf S|^m[\cos (m\theta_A)+\cos (m\theta_B)]$. The
  optimal choice of these degeneracy parameters is given in Table
  \ref{table1}}.
\end{thebibliography}

\end{document}